\documentclass[aps,print,showpacs,showkeys,10pt,twocolumn]{revtex4}
\usepackage{amssymb}
\usepackage{amsmath}
\usepackage{graphicx}

\input{tcilatex}
\begin{document}

\title{Screw-pitch effect and velocity oscillation of domain-wall in
ferromagnetic nanowire driven by spin-polarized current}
\author{Zai-Dong Li$^{1,2,3}$, Qiu-Yan Li$^{1}$, X. R. Wang$^{3}$, W. M. Liu$%
^{4}$, J. Q. Liang$^{5}$, and Guangsheng Fu$^{2}$}
\affiliation{$^{1}$Department of Applied Physics, Hebei University of Technology, Tianjin
300401, China\\
$^{2}$School of Information Engineering, Hebei University of Technology,
Tianjin, 300401, China\\
$^{3}$Physics Department, The Hong Kong University of Science and
Technology, Clear Water Bay, Hong Kong SAR, China.\\
$^{4}$Beijing National Laboratory for Condensed Matter Physics, Institute of
Physics, Chinese Academy of Sciences, Beijing 100080, China\\
$^{5}$Institute of Theoretical Physics and Department of Physics, Shanxi
University, Taiyuan 030006, China }

\begin{abstract}
We investigate the dynamics of domain wall in ferromagnetic nanowire with
spin-transfer torque. The critical current condition is obtained
analytically. Below the critical current, we get the static domain wall
solution which shows that the spin-polarized current can't drive domain wall
moving continuously. In this case, the spin-transfer torque plays both the
anti-precession and anti-damping roles, which counteracts not only the
spin-precession driven by the effective field but also Gilbert damping to
the moment. Above the critical value, the dynamics of domain wall exhibits
the novel screw-pitch effect characterized by the temporal oscillation of
domain wall velocity and width, respectively. Both the theoretical analysis
and numerical simulation demonstrate that this novel phenomenon arise from
the conjunctive action of Gilbert-damping and spin-transfer torque. We also
find that the roles of spin-transfer torque are entirely contrary for the
cases of below and above the critical current.
\end{abstract}

\maketitle

A magnetic domain wall (DW) is a spatially localized configuration of
magnetization in ferromagnet, in which the direction of magnetic moments
inverses gradually. When a spin-polarized electric current flows through DW,
the spin-polarization of conduction electrons can transfer spin momentum to
the local magnetization, thereby applying spin-transfer torque which can
manipulate the magnetic DW without the applied magnetic field. This
spin-transfer effect was theoretically proposed by Slonczewski \cite%
{Slonczewski} and Berger \cite{Berger}, and subsequently verified
experimentally \cite{Katine}. As a theoretical model the modified
Landau-Lifshitz-Gilbert (LLG) equation \cite{Bazaliy,Tatara,S.Zhang} with
spin-transfer torque was derived to describe such current-induced
magnetization dynamics in a fully polarized ferromagnet. With these novel
forms of spin torque many interesting phenomena have been studied, such as
spin wave excitation \cite{Bazaliy,Tsoi,Rezende} and instability \cite%
{Bazaliy,He}, magnetization switching and reversal \cite%
{Sun,Tsoi02,Chen,Jiang}, and magnetic solitons \cite{lizd,pbhe}. For the
smooth DW, this spin torque can displace DW opposite to the current
direction which has been confirmed experimentally in magnetic thin films and
magnetic wires \cite{Koo,Yama,Sai,Lim,Ohe}.

With the remarkable experimental success measuring the motion of DW under
the influence of current pulse, considerable progress has been made to
understand the current-induced DW motion in magnetic nanowire \cite%
{Tatara,S.Zhang,Yama,Sai,Lim,Ohe}. These studies have improved the
pioneering work of current-driven DW motion by Berger \cite{Berger2}.
Although both the theory and the quasi-static experiments have indicated
that the spin-polarized current can cause DW motion, the current-driven DW
dynamics is not well understood. The dynamics of magnetization described by
the LLG equation admits the static solutions for DW motion. In the presence
of spin torque and the external magnetic field, it is difficult to derive
the dynamic solutions. A circumvented approach is Walker solution analysis 
\cite{Walker} for the moving DW in response to a steady magnetic field
smaller than some critical value. However, this approximation applying to DW
motion driven by the electric current is unclear, and its reliability has to
be verified theoretically and numerically.

In this paper, we report analytically the critical current condition for
anisotropic ferromagnetic nanowire driven only by spin-transfer torque.
Below the critical current, the ferromagnetic nanowire admits only the final
static DW solution which implies that the spin-polarized current can't drive
DW moving continuously. When the spin-polarized current exceeds the critical
value, the dynamics of DW exhibits the \textit{novel Screw-pitch effect with
the periodic temporal oscillation of DW velocity and width}. A detail
theoretical analysis and numerical simulation demonstrate that this novel
phenomenon arises from the natural conjunction action of Gilbert-damping and
spin-transfer torque. We also observe that the spin-transfer torque plays
the entirely opposite roles in the above two cases. At last, our theoretical
prediction can be confirmed by the numerical simulation in terms of RKMK
method \cite{Munthe}.

We consider an infinite long uniaxial anisotropic ferromagnetic nanowire,
where the electronic current flows along the long length of the wire defined
as $x$ direction which is also the easy axis of anisotropy ferromagnet. For
convenience the magnetization is assumed to be nonuniform only in the
direction of current. Since the magnetization varies slowly in space, it is
reasonable to take the adiabatic limit. Then the dynamics of the localized
magnetization can be described by the modified LLG equation with
spin-transfer torque%
\begin{equation}
\frac{\partial \mathbf{M}}{\partial t}=-\gamma \mathbf{M\times H}_{\text{eff}%
}+\frac{\alpha }{M_{s}}\mathbf{M}\times \frac{\partial \mathbf{M}}{\partial t%
}+b_{J}\frac{\partial \mathbf{M}}{\partial x},  \label{LL1}
\end{equation}%
where $\mathbf{M}\equiv \mathbf{M}\left( x,t\right) $ is the localized
magnetization, $\gamma $ is the gyromagnetic ratio, $\alpha $ is the damping
parameter, and $\mathbf{H}_{\text{eff}}$ represents the effective magnetic
field. The last term of Eq. (\ref{LL1}) denotes the spin-transfer torque,
where $b_{J}=Pj_{e}\mu _{B}/(eM_{s})$, $P$ is the spin polarization of the
current, $j_{e}$ is the electric current density and flows along the $x$
direction, $\mu _{B}$ is the Bohr magneton, $e$ is the magnitude of electron
charge, and $M_{s}$ is the saturation magnetization. For the uniaxial
ferromagnetic nanowire the effective field can be written as $\mathbf{H}_{%
\text{eff}}=\left( 2A/M_{s}^{2}\right) \partial ^{2}\mathbf{M/}\partial x^{2}%
\mathbf{+}H_{x}M_{x}/M_{s}\mathbf{e}_{x}-4\pi M_{z}\mathbf{e}_{z}$, where $A$
is the exchange constant, $H_{x}$ is the anisotropy field, and $\mathbf{e}%
_{i}$, $i=x,y,z,$ is the unit vector, respectively. Introducing the
normalized magnetization, i.e., $\mathbf{m=M/}M_{s}$, Eq. (\ref{LL1}) can be
simplified as the dimensionless form%
\begin{eqnarray}
\alpha _{1}\frac{\partial \mathbf{m}}{\partial t} &=&-\mathbf{m\times h}_{%
\text{eff}}-\alpha \mathbf{m}\times \left( \mathbf{m\times h}_{\text{eff}%
}\right)  \notag \\
&&+\alpha b_{1}\mathbf{m}\times \frac{\partial \mathbf{m}}{\partial x}+b_{1}%
\frac{\partial \mathbf{m}}{\partial x},  \label{LL2}
\end{eqnarray}%
where $\alpha _{1}=\left( 1+\alpha ^{2}\right) $ and $b_{1}=b_{J}t_{0}/l_{0}$%
. The time $t$ and space coordinate $x$ have been rescaled by the
characteristic time $t_{0}=1/\left( 16\pi \gamma M_{s}\right) $ and length $%
l_{0}=\sqrt{A/\left( 8\pi M_{s}^{2}\right) }$, respectively. The
dimensionless effective field becomes $\mathbf{h}_{\text{eff}}=\partial ^{2}%
\mathbf{m}/\partial x^{2}+C_{1}m_{x}\mathbf{e}_{x}-C_{2}m_{z}\mathbf{e}_{z}$%
, with $C_{1}=H_{x}/\left( 16\pi M_{s}\right) $ and $C_{2}=0.25$.

In the following, we seek for the exact DW solutions of Eq. (\ref{LL2}), and
then study the dynamics of magnetization driven by spin-transfer torque. To
this purpose we make the ansatz 
\begin{equation}
m_{x}=\tanh \Theta _{1},m_{y}=\frac{\sin \! \phi }{\, \cosh \Theta _{1}}%
,m_{z}=\frac{\cos \! \phi }{\cosh \Theta _{1}},  \label{DW1}
\end{equation}%
where $\Theta _{1}=k_{1}x-\omega _{1}t$, with the temporal and spatial
independent parameters $\! \phi $, $k_{1}$, and $\omega _{1}$ to be
determined, respectively. Substituting Eq. (\ref{DW1}) into Eq. (\ref{LL2})
we have%
\begin{eqnarray}
k_{1}^{2} &=&C_{1}+C_{2}\cos ^{2}\phi ,  \label{so2a} \\
-\omega _{1}\left( 1+\alpha ^{2}\right) &=&b_{1}k_{1}+C_{2}\sin \phi \cos
\phi ,  \label{so2b} \\
\alpha b_{1}k_{1}\cos \phi &=&\alpha \left( C_{1}-k_{1}^{2}\right) \sin \phi
,  \label{so2c} \\
\alpha b_{1}k_{1}\sin \phi &=&-\alpha C_{2}\sin ^{2}\phi \cos \phi .
\label{so2d}
\end{eqnarray}%
From the above equations we can get three cases of DW solutions for Eq. (\ref%
{LL2}). Firstly, in the absence of damping Eqs. (\ref{so2a}) to (\ref{so2d})
admit the solution 
\begin{equation}
k_{1}=\pm \sqrt{C_{1}+C_{2}\cos ^{2}\phi },\omega _{1}=-b_{1}k_{1}-\frac{%
C_{2}}{2}\sin 2\phi ,  \label{para1a}
\end{equation}%
with the arbitrary angle $\phi $. This solution show that the spin-transfer
torque contributes a dimensionless velocity $-b_{1}$ only without damping.
The velocity of DW is formed by two parts, i.e., $v=-\left( C_{2}\sin 2\phi
\right) /\left( 2k_{1}\right) -b_{1}$, which can be affected by adjusting
the angle $\phi $ and the spin-transfer torque. Secondly, in the absence of
spin torque, we have the solution of Eqs. (\ref{so2a}) to (\ref{so2d}) as $%
\omega _{1}=0$, $\phi =\pm \pi /2$, $k_{1}=\pm \sqrt{C_{1}}$, i.e., the
static DW solution. In terms of RKMK method \cite{Munthe} we perform direct
numerical simulation for Eq. (\ref{LL2}) with various initial condition, and
all numerical results show that the damping drives the change of $\phi $
which in turn affects the DW velocity and width defined by $1/\left \vert
k_{1}\right \vert $. At last $\phi =\pm \pi /2,\omega _{1}=0$, i.e., the DW
loses moving, and the DW width attains its maximum value $\sqrt{C_{1}}$,
which confirms the Walker's analysis \cite{Walker} that the damping prevents
DW from moving without the external magnetic field or spin-transfer torque.
However, as shown later, the presence of damping is prerequisite for the
novel \textit{Screw-pitch }property of DW driven by spin-transfer torque. At
last, we consider the case of the presence of damping and spin-transfer
torque. Solving Eqs. (\ref{so2a}) to (\ref{so2d}) we have 
\begin{equation}
k_{1}=\pm \frac{1}{2}(B_{1}-\sqrt{B_{2}}),\omega _{1}=0,\sin 2\phi =-\frac{%
2b_{1}k_{1}}{C_{2}},  \label{para1}
\end{equation}%
where $B_{1}=2C_{1}+C_{2}-b_{1}^{2}$, $B_{2}=\left( C_{2}-b_{1}^{2}\right)
^{2}-4C_{1}b_{1}^{2}$.

It is clear that Eq. (\ref{para1}) implies the critical spin-polarized
current condition, namely 
\begin{equation*}
b_{J}\leq (\sqrt{C_{1}+C_{2}}-\sqrt{C_{1}})l_{0}/t_{0},
\end{equation*}%
which is determined by the character velocity $l_{0}/t_{0}$, the anisotropic
parameter $C_{1}$, and the demagnetization parameter $C_{2}$. Below the
critical current, i.e., $b_{1}^{2}\leq (\sqrt{C_{1}+C_{2}}-\sqrt{C_{1}})^{2}$%
, the DW width falls into the range that $1/\sqrt{C_{1}^{2}+C_{1}C_{2}}\leq
1/\left \vert k\!_{1}\right \vert \leq 1/C_{1}$. From Eq. (\ref{para1}) we
get four solutions of $\phi $, i.e., $\phi =\pm \pi /2+1/2\arcsin \left(
2b_{1}k_{1}/C_{2}\right) $ for $k\!_{1}>0$ and $\phi =\pm \pi /2-1/2\arcsin
\left( 2b_{1}\left \vert k_{1}\right \vert /C_{2}\right) $ for $k\!_{1}<0$.
In fact, the signs \textquotedblleft $+$\textquotedblright \ and
\textquotedblleft $-$\textquotedblright \ in Eq. (\ref{para1}) denotes kink
and anti-kink solution, respectively, and the corresponding solution in Eq. (%
\ref{DW1}) represents the static tail-to-tail or head-to-head N\'{e}el DW,
respectively. This result shows that below the critical current, the final
equilibrium DW solution must be realized by the condition that $\mathbf{%
m\times h}_{\text{eff}}=b_{1}\partial \mathbf{m}/\partial x$. It clearly
demonstrates that the spin-transfer torque has two interesting effects. One
is that the term $b_{1}\partial \mathbf{m}/\partial x$ in Eq. (\ref{LL2})
plays the anti-precession role counteracting the precession driven by the
effective field $\mathbf{h}_{\text{eff}}$. However, the third term in the
right hand of Eq. (\ref{LL2}), namely $\alpha b_{1}\mathbf{m}\times \partial 
\mathbf{m}/\partial x$, has the anti-damping effect counteracting the
damping term $-\alpha \mathbf{m}\times (\mathbf{m\times h}_{\text{eff}})$.
It is to say that below the critical value, the spin-polarized current can't
drive DW moving continuously without the applied external magnetic field.

When the spin-polarized current exceed the critical value, the dynamics of
DW possesses two novel properties as shown in the following section. Above
the critical current, the precession term $-\mathbf{m\times h}_{\text{eff}}$
can't be counteracted by spin-transfer torque, and the static DW solution of
Eq. (\ref{LL2}) doesn't exist. Because the magnetization magnitude is
constant, i.e., $\mathbf{m}^{2}=1$, so we have $\mathbf{m}\cdot \partial 
\mathbf{m}/\partial x=0$ which shows that the direction of $\partial \mathbf{%
m}/\partial x$ is always perpendicular to the direction of $\mathbf{m}$, or $%
\partial \mathbf{m}/\partial x=0$. It is well known that a magnetic DW
separates two opposite domains by minimizing the energy. In the magnetic DW
the direction of magnetic moments gradually changes, i.e., $\partial \mathbf{%
m}/\partial x\neq 0,$ so the direction of $\partial \mathbf{m}/\partial x$
should adopt the former case. Out of region of DW the normalized
magnetization will site at the easy axis, i.e., $m_{x}=1$(or $-1$), in which 
$\mathbf{\partial m}/\partial x=0$.

With the above consideration we make a detail analysis for Eq. (\ref{LL2}).
As a characteristic view we mainly consider the DW center, defined by $%
m_{x}=0$. The magnetic moment must be in the $y$-$z$ plane, while the
direction of $\partial \mathbf{m}/\partial x$ should lay in $x$-axis ($+x$%
-axis for $k_{1}>0$, and $-x$-axis for $k_{1}<0$). In order to satisfy Eq. (%
\ref{LL2}) the magnetic moment in DW center should include both the
precession around the effective spin-torque field $\alpha b_{1}\partial 
\mathbf{m}/\partial x$ and the tendency along the direction of $\partial 
\mathbf{m}/\partial x$ continuously from the last two terms in the right
hand of Eq. (\ref{LL2}). The former precession motion implies that the
parameter $\phi $ will rotate around $x$-axis continuously, while the latter
tendency forces the DW center moving toward to the opposite direction of the
current, i.e., $-x$-axis direction, confirming the experiment \cite%
{Koo,Yama,Sai,Lim,Ohe} in magnetic thin films and magnetic wires. Combining
the above two effects we find that this rotating and moving phenomenon is
very similar to \textit{Screw-pitch effect. }The continuous rotation of
magnetic moment in DW center, i.e., the periodic change of $\phi $, can
result in the periodic oscillation of DW velocity and width from Eq. (\ref%
{para1a}) under the action of the first two terms in the right hand of Eq. (%
\ref{LL2}). It is interesting to emphasize that when the current exceeds the
critical value, the term $\alpha b_{1}\mathbf{m}\times \partial \mathbf{m}%
/\partial x$ plays the role to induce the precession, while the term $%
b_{1}\partial \mathbf{m}/\partial x$ has the effect of damping, which is
even entirely contrary to the case below the critical current as mentioned
before. Combining the above discussion we conclude that the motion of
magnetic moment in the DW center will not stop, except it falls into the
easy axis, i.e., out of the range of DW. In fact, all the magnetic moments
in DW can be analyzed in detail with the above similar procedure.

Now it is clear for the dynamics of DW driven only by spin-transfer torque.
Coming back to Eq. (\ref{LL2}) we can see that this novel \textit{%
Screw-pitch effect }with the periodic oscillation of DW velocity and width
occurs even at the conjunct action of the damping and spin-transfer torque.
To confirm our theoretical prediction we perform direct numerical simulation
for Eq. (\ref{LL2}) with an arbitrary initial condition by means of RKMK
method \cite{Munthe} with the current exceeding the critical value. In
figure 1(a) to 1(c) we plot the time-evolution of the normalized
magnetization $\mathbf{m}$, while the displacement of DW center is shown in
figure 1(d). The result in figure 1 confirms entirely our theoretical
analysis above. The evolution of $\cos \phi $ and the DW velocity and width
are shown in figure 2. From figure 2 we can see that the periodic change of $%
\cos \phi $ leads to the periodic temporal oscillation of DW velocity and
width. From Eq. (\ref{para1a}) and the third term of Eq. (\ref{LL2}) we can
infer that $\cos \phi $ possesses of the uneven change as shown in figure
2(a), i.e., the time corresponding to $0<\phi +n\pi \leq \pi /2$ is shorter
than that corresponding to $\pi /2<\phi +n\pi \leq \pi $, $n=1,2...$, in
each period, and the DW velocity has the same character. It leads to the DW
displacement firstly increases rapidly, and then slowly as shown in figure
1(d). This phenomenon clarifies clearly the presence of \textit{Screw-pitch
effect}.\textit{\ }The DW velocity oscillation driven by the external
magnetic field has been observed experimentally \cite{Bea}. Our theoretical
prediction for the range of DW velocity oscillation driven by the above
critical current could be observed experimentally.

In summary, the dynamics of DW in ferromagnetic nanowire driven only by
spin-transfer torque is theoretically investigated. We obtain an analytical
critical current condition, below which the spin-polarized current can't
drive DW moving continuously and the final DW solution is static. An
external magnetic field should be applied in order to drive DW motion. We
also find that the spin-transfer torque counteracts both the precession
driven by the effective field and the Gilbert damping term different from
the common understanding. When the spin current exceeds the critical value,
the conjunctive action of Gilbert-damping and spin-transfer torque leads
naturally the novel screw-pitch effect characterized by the temporal
oscillation of DW velocity and width.

This work was supported by the Hundred Innovation Talents Supporting Project
of Hebei Province of China, the NSF of China under grants Nos 10874038,
10775091, 90406017, and 60525417, NKBRSFC under grant No 2006CB921400, and
RGC/CERG grant No 603007.

\textbf{Figure Captions}\newline

Fig. 1. The dynamics of DW above the critical current. (a)-(c) Evolution of
the normalized magnetization $\mathbf{m}$. (d) The displacement of DW driven
only by spin-transfer torque. The parameters are $\alpha =0.2,$ $%
C_{1}=0.05,C_{2}=0.25,b_{J}=0.6$, and the initial angle $\phi =0.01\pi $.

Fig. 2. (a) The evolution of $\cos \phi $ and the periodic oscillation of DW
velocity. (b) The periodic temporal oscillation of DW width. The parameters
are same as in figure 1.

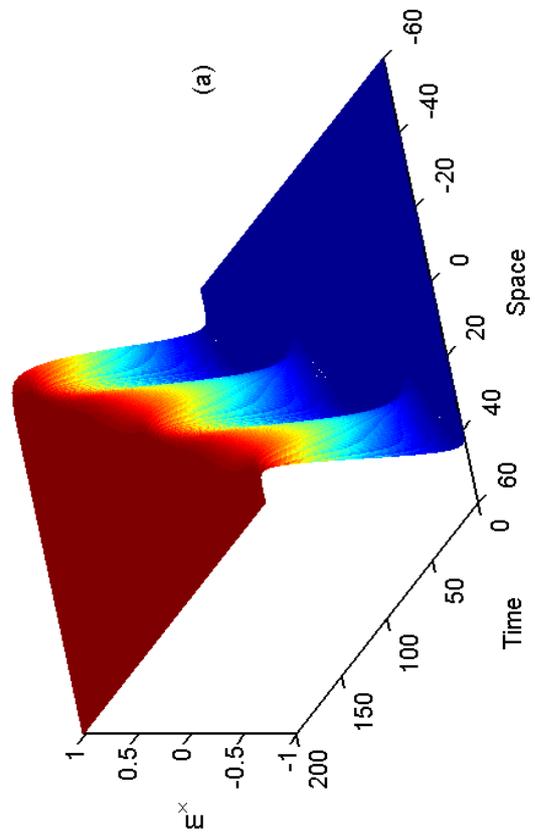
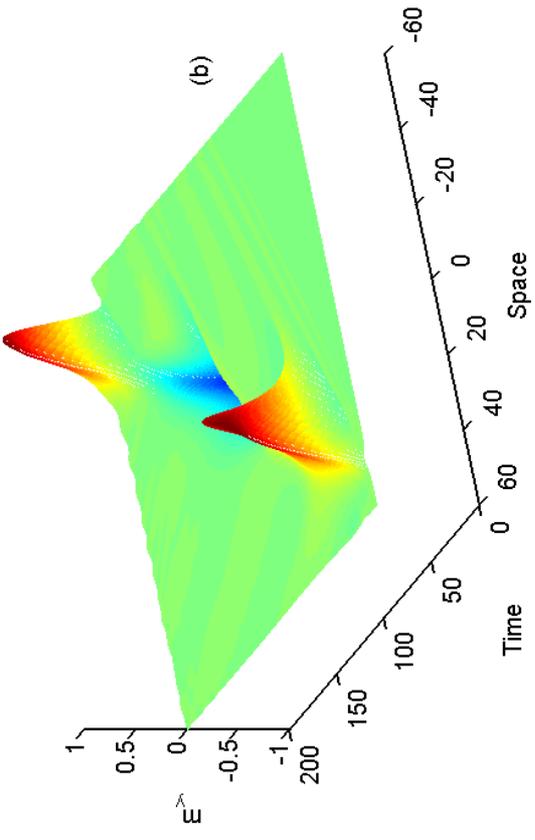
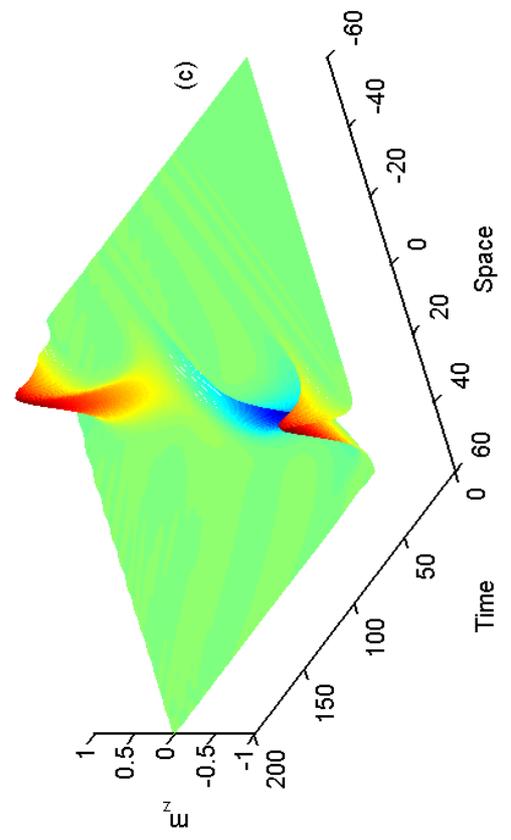
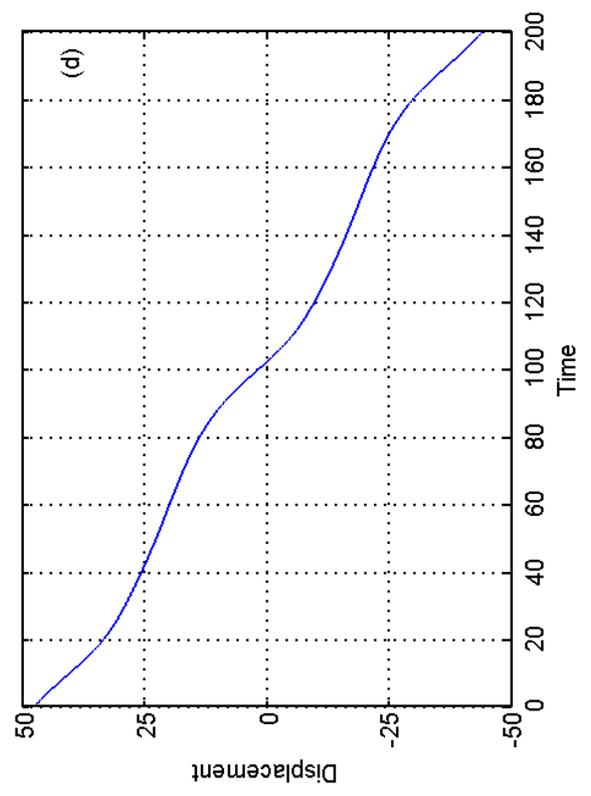
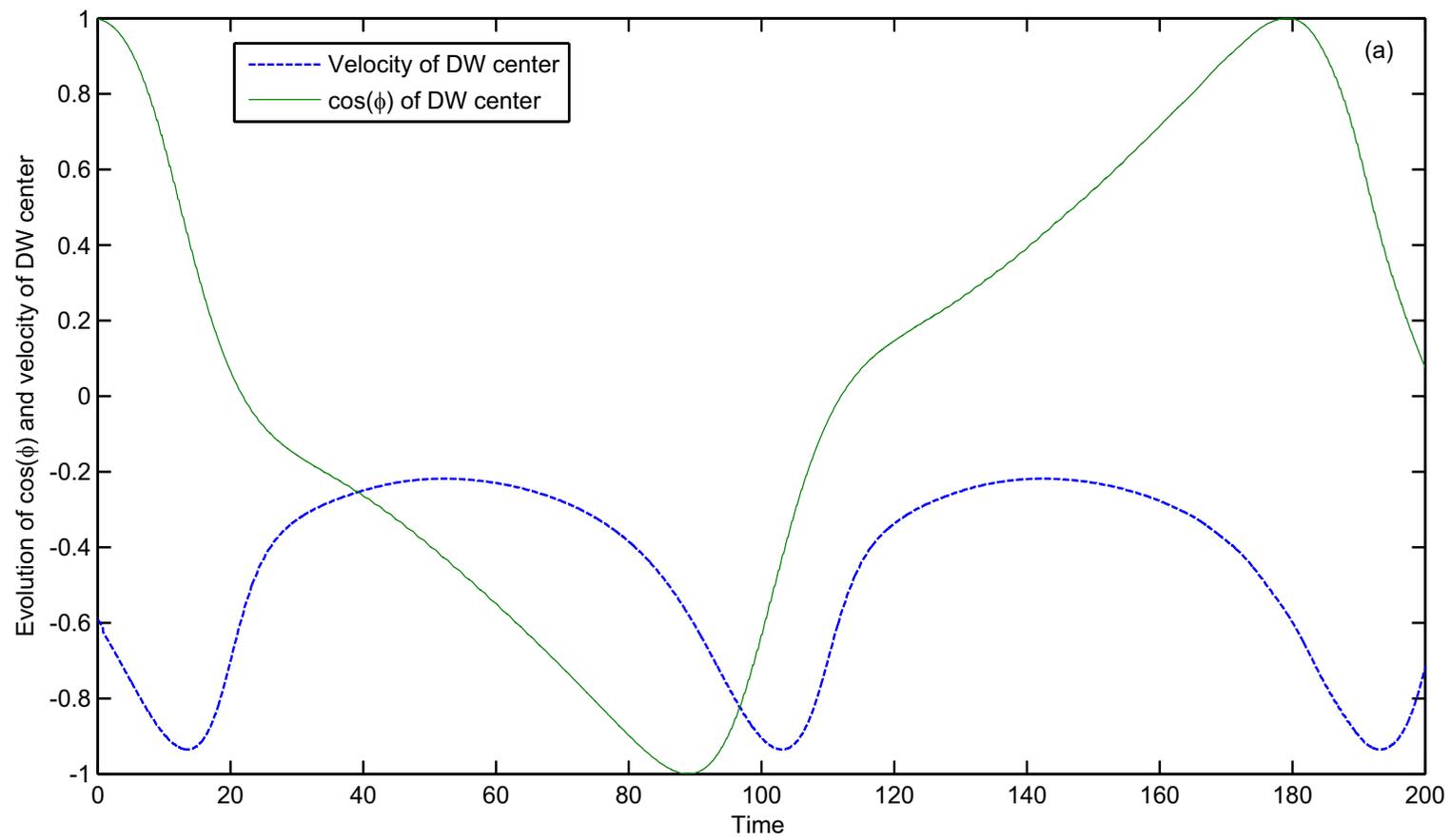
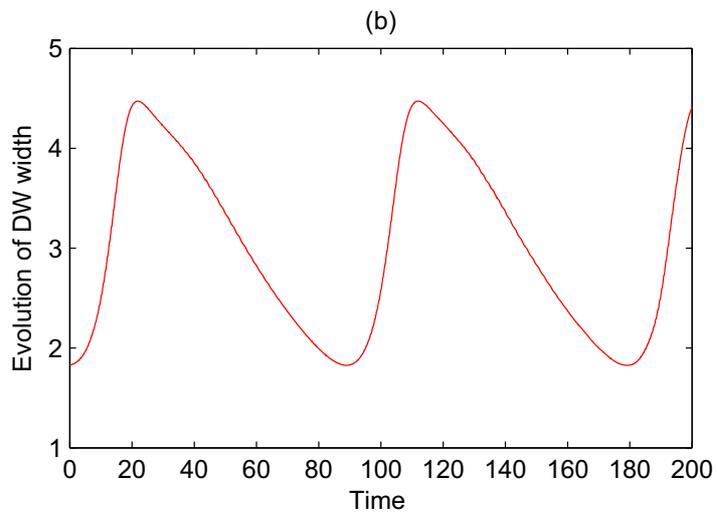
\end{document}